\documentstyle[aas2pp4]{article}

\def\deg{\hbox{$^\circ$}}
\def\arcmin{\hbox{$^\prime$}}
\def\arcsec{\hbox{$^{\prime\prime}$}}

\def\aspt{\rlap{.}$''$}
\def\kms{km\,s$^{-1}$}
\def\ha{H$\alpha$}
\def\rosat{{\em ROSAT}}

\begin{document}

\title{Hot Interstellar Gas in the Irregular Galaxy NGC\,4449\footnotemark}
\author{Dominik J. Bomans\altaffilmark{2} and You-Hua Chu}
\affil{Astronomy Department, University of Illinois, 1002 W. Green St., 
Urbana, IL 61801, USA}

\and

\author{Ulrich Hopp\altaffilmark{2}}
\affil{Universit\"atssternwarte M\"unchen, Scheiner Str. 1, 81679 
M\"unchen, Germany}

\footnotetext{Partly based on observations made with the NASA/ESA Hubble 
Space Telescope, obtained 
from the data archive at the Space Telescope Science Institute. STScI is 
operated by the Association of Universities for Research in Astronomy, Inc. 
under the NASA contract NAS 5-26555.}
\altaffiltext{2}{Feodor Lynen Fellow of the Alexander von Humboldt-Foundation} 
\altaffiltext{3}{Visiting Astronomer, Calar Alto Observatory}

\begin{abstract}

NGC 4449 is an irregular galaxy with a moderately high star formation 
activity.  The massive stars in NGC 4449 have given rise to many 
bright HII regions, superbubbles, supergiant shells, and
``chimney-like" radial filaments.  {\em ROSAT} X-ray observations of
NGC\,4449 have revealed four point-like sources and a 
wide-spread diffuse emission.  The spectral properties of the diffuse 
component suggest that the emission originates from hot interstellar 
gas.  We have compared deep ground-based \ha\ images with the X-ray 
images of NGC\,4449 to determine the relationship between the hot 
(10$^6$ K) and the warm (10$^4$ K) components of the interstellar gas.
We have also used an archival {\em Hubble Space Telescope} WFPC2 image 
of NGC\,4449 taken through the F606W filter to examine the massive stellar 
content of the X-ray-emission regions.  We find that hot interstellar gas 
exists in 
(1) active star forming regions, including the giant H\,II region CM\,16, 
(2) probable outflows from star forming regions, and (3) the supergiant shell 
SGS2.  The X-ray data have been used to derive the rms electron 
density, mass, and thermal energy of the hot interior of SGS2.
Finally we discuss the origin of SGS2 and implications of the detection 
of diffuse X-rays in irregular galaxies.

\end{abstract}


\newpage

\section{Introduction}

Diffuse X-ray emission has been detected in late-type galaxies, 
indicating the existence of hot (10$^6$ K) interstellar gas 
(e.g., M33 -- \cite{Long96}; M101 -- \cite{Snowden95}).
The hot interstellar gas can be studied with particularly high 
spatial resolution in the Large and Small Magellanic Clouds (LMC 
and SMC), as these are the two nearest neighboring galaxies.  The 
{\em ROSAT} observations of the LMC has shown unprecedented details 
of the bright diffuse X-ray emission associated with the giant H\,II 
region 30 Doradus complex, large shell structures, as well as 
regions with no obvious ionized interstellar structures (\cite{Snowden94}; 
\cite{Chu96}; \cite{Bomans97}).  The SMC, on the other 
hand, shows very little diffuse X-ray emission (\cite{Snowden96}).

What determines the level of diffuse X-ray emission, or the amount
of hot interstellar gas?  Is the bright diffuse emission of the 
LMC or the lack of diffuse emission in the SMC the norm for 
irregular galaxies?   To answer these questions, other irregular
galaxies need to be studied in detail.  We have chosen the irregular 
galaxy NGC\,4449, which is similar to the LMC in size and mass 
(\cite{Bajaja94}) and has been relatively well-studied in many
wavelengths.

As a probable member of the Canis Venaticorum galaxy group, NGC\,4449
is at a distance of 3 Mpc (\cite{Tully88}) or 5.4 Mpc (\cite{Schmidt92}); 
we adopt the latter.  The inclination of NGC\,4449 determined from 
isophote analysis of the continuum light
is $\sim$43\deg\ (\cite{Tully88}).   NGC\,4449 is one of the first 
galaxies in which diffuse ionized filaments were detected 
(\cite{Sabbadin79}).  Its moderately high star formation activity 
apparently has produced long filaments and a frothy structure of the warm 
(10$^4$ K) ionized interstellar medium (\cite{Hunter90}; \cite{Hunter92}; 
hereafter HG90, HG92).  
An H\,I halo, extending over more than 1\deg, is detected around
NGC\,4449 (\cite{vanWoerden75}; \cite{Bajaja94}).  Recently, Klein et al. 
(1996) detected a large synchrotron emission halo around 
NGC\,4449, and speculated the existence of a hot gaseous halo.  They 
also found an ordered magnetic field on kpc scales, the first detection
in an irregular galaxy other than the LMC. 
A super-luminous supernova remnant (SNR) in NGC\,4449 has been detected 
in both radio continuum (\cite{Seaquist78}) and X-ray 
(\cite{Blair83}).

X-ray point sources in NGC\,4449 have been detected by {\em Einstein}
observations; however, the diffuse emission in NGC\,4449 is not obvious until
{\em ROSAT} observations become available.  
In this paper we report our analysis of X-ray and optical observations 
of the interstellar medium in NGC\,4449.  The observations and data 
reduction are described in \S 2.  We discuss the large interstellar 
shell structures and filaments of NGC\,4449 in \S 3.  The nature of 
the X-ray sources and the relationship between the hot and warm 
interstellar gases are analyzed in \S 4.  Finally, in \S 5 we
discuss the relationship between the 10$^6$ K hot gas and massive 
stars, examine the hot interior of the supergiant shell SGS2, and discuss 
the origin of SGS2 and implications of the detection 
of diffuse X-rays in irregular galaxies.

\section{Observations and Reductions}

\subsection{Optical Observations}

To examine the warm ionized interstellar gas, two sets of optical 
emission-line and continuum images were obtained.  The primary set of 
images, having a large field of view and high sensitivity but taken 
under non-photometric conditions, are used to examine the global
structure.  The secondary set of images, having a smaller field of 
view but multiple wavelength coverage, are used to flux-calibrate 
the primary data and to investigate the ionization and excitation 
condition of the H\,II gas.

The primary set of images of NGC\,4449 were obtained with the Calar 
Alto Observatory 3.5m telescope in January 1991. The prime focus focal 
reducer was used, giving an effective focal ratio of 1:2.7.  The 
detector was a GEC 1152$\times$768 CCD with a pixel size of 0\aspt495 and 
a readout noise of 8.4 electrons.  Images were taken through a 100 \AA\ 
wide \ha\ filter and a 180 \AA\ wide red continuum filter.  
(Note that the H$\alpha$ filter transmits the [N II]$\lambda\lambda$6548, 
6583 lines as well.)  The seeing 
was 1\aspt5 during the observations.  The weather conditions were 
not stable during the night, which prohibited a direct flux calibration. 
The data were reduced following a standard procedure.  Because of the 
vignetting of the focal reducer, we have trimmed the 
final images to 920$\times$766 arrays, which correspond to about 
7\farcm6 $\times$ 6\farcm3.

The continuum contribution to the \ha\ image was removed by subtracting the
registered continuum image. Flux scaling between the images was done using 
isolated stars visible in both frames outside the main body of NGC\,4449 
and allowing for small adjustments to optimize the subtraction. 
The point spread functions (PSFs) of the on- and off-line images were so 
different that relatively large residuals remained after the subtraction. 
We therefore convolved both frames with narrow Gaussian filters to minimize
the differences in the PSF, withstanding a small trade-off in resolution. 
Continuum subtraction using these frames left clearly smaller and
more symmetric residuals. The resolution of these images was 2\arcsec.
The high surface brightness of NGC\,4449 further complicated the 
determination of an optimal scaling factor.
We decided to use a factor which might lead to a slight over-subtraction 
(visible at the eastern edge of the bar of NGC\,4449), in order to ensure 
that the resultant line emission did not contain residual stellar continuum.

The continuum image of NGC\,4449 is shown in Fig.\ \ref{fig1}, the 
continuum-subtracted \ha\ image of NGC\,4449 is shown in Fig.\ \ref{fig2}.  
In Fig.\ \ref{fig3} 
the central portion of the continuum-subtracted \ha\ image is overlaid 
with sketches of the principal filamentary nebular features and 
reference numbers to ease later discussions.  Positions and remarks of 
these features are tabulated in the Appendix, Table \ref{tabA1} for shells and 
Table \ref{tabA2} for irregular filaments.
The faint emission blob marked with ``ghost'' in Fig.\ \ref{fig3} is an 
internal reflection of the neighboring bright stars produced in the optical 
path of the instrument.

The secondary data set of NGC\,4449 was obtained with the Calar Alto 3.5m
telescope in March 1988 using the normal prime focus CCD camera equipped 
with a 580$\times$380 GEC chip. 
The field of view was 4\arcmin$\times$3\arcmin. Observations were made 
through filters centered at H$\alpha$, H$\beta$, [OIII], red continuum 
at 870 nm, and Johnson V, respectively. The seeing 
was 2\arcsec, but the sky was photometric. Planetary nebulae from 
the list of Kaler (\cite{Kaler76}; \cite{Kaler83}) were observed as emission 
line standards. 

The flux-calibrated secondary images provide an independent flux 
calibration for our primary \ha\ image.  We use a small, reasonably 
isolated H\,II region at 12$^{\rm h}28^{\rm m}08^{\rm s}$,
+44{\deg}06\farcm25 (J2000) for the flux transformation.  
Using the filter curves and an [NII]$\lambda$6583/\ha\ 
ratio of 0.15 for the ionized gas in NGC\,4449 (\cite{Lequeux79}), we 
estimate the contribution of the [NII] line to the primary image to be 10\%. 
The resultant limiting surface brightness of our primary \ha\ image is 
$9.0 \times 10^{-18}$ erg cm$^{-2}$ s$^{-1}$ arcsec$^{-2}$, corresponding
to a limiting emission measure of 4.5 cm$^{-6}$ pc if the electron
temperature is 10$^4$ K. 
This limiting emission measure could be still lower for the diffuse ionized 
gas, since the [NII] lines can be enhanced in the case of dilution of the 
ionizing photon field (e.g. \cite{Domgorgen94}).

\subsection{X-ray Observations}

NGC\,4449 has been observed by the {\em Einstein Observatory} with an
Imaging Proportional Counter (IPC; I2123, 1.7 ks) and a High Resolution 
Imager (EHRI; H4967, 32.4 ks), and the {\em R\"ontgensatellit (ROSAT)} 
with a Position Sensitive Proportional Counter (PSPC; RP600137, 
7.85 ks) and a High Resolution Imager (RHRI; WH600743, 20.1 ks;
WH600865, 41.3 ks).  The \rosat\ PSPC observation, having the highest 
sensitivity, is used to extract spectral information and to analyze diffuse
emission.  Both EHRI and RHRI observations are used to distinguish point
sources from the diffuse emission.

The X-ray observations of NGC\,4449 were retrieved from the High Energy 
Astrophysics Science Archive Research Center (HEASARC) at Goddard Space
Flight Center of NASA.  The software packages of IRAF/PROS\footnote{IRAF
is distributed by the National Optical Astronomy Observatories, which is 
operated by AURA, Inc. under cooperative agreement with the NSF. PROS is 
developed, distributed, and maintained by the Smithsonian Astrophysical 
Observatory, under partial support from NASA contracts NAS5-30934 and 
NAS8-30751.} were used for the data analysis.  
The PSPC image of NGC\,4449 in the 0.1--2.4 keV band smoothed with a
Gaussian of $\sigma$=10$''$ is displayed in Fig.\ \ref{fig4}a; the RHRI image 
WH600743 smoothed with a Gaussian of $\sigma$=3$''$ is displayed in
Fig.\ \ref{fig4}b.

The X-ray images of NGC\,4449 show point sources as well as diffuse 
emission.  The X-ray emission in the PSPC image can be divided
into seven discrete regions.  These regions are marked in Fig.\ \ref{fig5}.
The EHRI and RHRI images of NGC\,4449 show that regions 1, 2, and 3 
are dominated by bright point sources, and region 5 contains a weaker
point source superimposed on diffuse emission.  Note that the diffuse
emission in the northern part of NGC\,4449, centered on region 6,
is detected even in the RHRI image.

To align the X-ray and optical images of NGC\,4449, we use the
super-luminous SNR whose position is well determined from radio and optical 
observations (\cite{Seaquist78}).  This SNR corresponds to our 
X-ray source 2 in Fig.\ \ref{fig5}, and its optical counterpart is marked in 
Fig.\ \ref{fig3}.  The alignment between the X-ray and optical images should 
be accurate to about 2$''$.
In Fig.\ \ref{fig6}a we show a grey-scale \ha\ image of NGC\,4449 overlaid 
by X-ray contours derived from the PSPC data, while Fig.\ \ref{fig6}b 
overlayed by X-ray contours from the HRI data.

We have extracted background-subtracted spectra for the seven regions 
marked in Fig.\ \ref{fig5} and a region (No.\ 8) that includes the entire
galaxy.  The background is scaled from regions located outside but 
close to the galaxy.  Two different background regions have been 
used, but the resultant spectra and fluxes are virtually identical
within error limits.  For sources with sufficient counts, we have 
fitted the spectra using the Raymond \& Smith (1977) thin plasma 
emission model with a 30\% solar abundance, as measured from
H\,II regions in NGC\,4449 (\cite{Lequeux79}).
For sources with insufficient counts ($<$150 counts) for spectral fits,
we adopt the absorption column density determined from the spectral fits
of the brightest sources, and estimate plasma temperatures based on the
spectral shape.  We compare the spectral shape to those of other 
plasma emission sources, such as giant H\,II regions in M101 
(\cite{Williams95}) and superbubbles and supergiant shells in the LMC 
(\cite{Chu93}; \cite{Bomans94}), and adopt the corresponding plasma 
temperatures.
The number of source counts, foreground absorption column densities, 
plasma temperatures, and X-ray luminosities in the 0.1--2.4 keV and
0.5--2.4 keV bands of the seven regions and of the whole galaxy are
tabulated in Table \ref{tab1}.

For a quantitative analysis of the spectral properties of the apparent 
diffuse emission a subtraction of the three point source would be 
extremely useful. We did not apply this step for the following reasons:
first, the point-sources are relatively close to each other and no isolated 
bright point-source is available as a template, therefore the exact 
subtraction of the wings of the point-spread function is very uncertain. 
Secondly, this exact subtraction is critical since the underlying possible 
diffuse emission in the area of the galaxy bar is faint. 
We used instead images in selected energy bands to investigate 
the large scale distribution of the diffuse emission in NGC\,4449 (see 
section 4.2).  Observations with the forthcoming X-ray satellites will 
provide the needed spatial and spectral resolutions.

\section{Large Interstellar Shells and Filaments in NGC\,4449}

Star formation is quite active in NGC\,4449.  Nearly the entire
optical extent of the galaxy is covered with H\,II regions and prominent
filamentary structures.  The two most luminous H\,II regions, CM\,16 and
CM\,39 (\cite{Crillon69}), rival the archetypical giant H\,II 
region 30\,Dor in both size and luminosity.  
The filamentary structures have been previously detected and discussed by 
HG90, HG92, Bomans \& Hopp (1993), and Hill et al. (1994).  We have 
identified the most prominent filaments and marked them in Fig.\ \ref{fig3}. 
Some of the filaments form large shell structures, such as supergiant shells 
and superbubbles (listed in Table \ref{tabA1}), some filaments extend 
radially, while others are irregular or multiple (listed in 
Table \ref{tabA2}).  
Many filaments may be physically associated with  large star formation 
regions.   For example, the giant H\,II region CM\,16 appears to be 
connected with the superbubbles SB3 and SB4 and the filaments FIL27--29 
and perhaps FIL6 and FIL30, forming a complex structure covering 1 kpc.
Below we discuss supergiant shells, superbubbles, and other 
filaments separately.

We have identified five supergiant shells with diameter greater than 
500 pc, SGS1--5.  SGS1 was previously identified by HG90.  SGS2 
corresponds to HG92's filaments 5 and 6.  SGS3--5 have not been 
previously documented.  Note that our division between supergiant shells
and superbubbles may not be physical.  The largest superbubbles have 
sizes approaching the 500 pc threshold for supergiant shells.  
Furthermore, if NGC\,4449 is at a smaller distance (3 Mpc, \cite{Tully88}), 
then SGS1 and SGS3 will not be qualified as supergiant shells.

The five supergiant shells have very different physical properties. 
See Fig.\ \ref{fig3} for the outlines of these supergiant shells.  
SGS1 is adjacent to the giant H\,II region CM\,16, reminiscent of the
supergiant shell LMC2 to the giant H\,II region 30 Dor in the LMC.  
It contains a luminous OB association.  SGS2 consists of a set of 
very long ($>1$ kpc) curved filaments west of the ridge of intense 
star formation. The mid section of SGS2 lies at distances up to 
2 kpc from the star formation regions.  
SGS2 is the largest known ionized supergiant shell.  SGS3, like SGS1, 
encompasses a luminous OB association; however, unlike the other 
supergiant shells, SGS3 is isolated at 3 kpc away from the high surface 
brightness bar of the galaxy.  SGS4 is faint, with no apparent concentration 
of luminous associations.  SGS5 is faint and is superimposed on a complex
background; its identification may be uncertain.

We use ``superbubbles'' to refer to large shells with diameters $<$500 pc.
We have identified nine superbubbles, SB1--9.
Three superbubbles, SB3, SB4 and SB5, appear to be ``blisters'' 
extending from giant H\,II regions.  The others are associated with OB
associations, similar to the common superbubbles in the LMC, such
as N44 (\cite{Oey95}; \cite{Will97}).

The non-shell type filamentary structures we cataloged are by no means
complete, because crowding in the main body of the galaxy prohibits reliable 
identifications of nebular structures.  Some of the filaments have been 
previously identified by HG90; these are noted in Table \ref{tabA2}.  
Among these non-shell type filaments, the most remarkable ones are those 
oriented perpendicular to the major axis of NGC\,4449 -- FIL3, 15, 16, 19, 
22, 23, 24, 25, 26, 27, 28, 29, and 37.  These filaments are reminiscent of 
the chimneys in edge-on galaxies such as NGC\,891 (\cite{Dettmar90}; 
\cite{Rand90}) and possibly the H\,I worms in our galaxy (\cite{Heiles84}).  
Interestingly, the distribution of filaments in NGC\,4449 is far from 
uniform; the western side of the galaxy is populated by numerous radial 
filaments, while the eastern side has fewer filaments and the filaments 
are tangential to the major axis of the galaxy.  This distribution 
suggests that energy and gas outflows into the halo take place 
preferentially on the western side of the galaxy.  This will be discussed 
further in \S 5.
The interpretation of radial filaments as chimneys is based on our 
understanding of dwarf irregular galaxies (see e.g. \cite{Skillman94}) 
that NGC\,4449 is a flattened body with a large scale height for its 
neutral gas.

Finally, we note that the [O III]/H$\alpha$ ratios of the brightest
filaments (e.g., FIL2 and FIL3) are lower than those of their adjacent
H\,II regions, regardless of their distance form the H\,II regions.  
Similar behavior of the [O III]/H$\alpha$ ratio has been observed 
in SGSs and filaments in the LMC (\cite{Hunter94}).  This seems to indicate 
that the filaments are not photo-ionized in the same way as the calssical 
HII regions by UV radiation of O stars.  
Other considerations need to be taken into account, such as dilution of the 
photon-field (\cite{Domgorgen94}), an intrinsically very soft ionizing 
photon field (e.g. \cite{Skillman97}), turbulent mixing layers 
(\cite{Slavin93}), or ionization by soft X-rays from cooling hot gas.

\section{Hot Gas in NGC\,4449}

Hot (10$^6$ K) interstellar gas is best studied in X-rays.  However, 
stellar coronal emission and X-ray binaries contribute to the X-ray
flux observed.  
Thus we need to consider carefully the nature of the
X-ray sources before we use the X-ray emission to extract information 
on the hot interstellar gas.  
Note that individual SNRs in NGC\,4449, if detected, would appear as 
point sources.

\subsection{Physical Nature of the X-ray Sources}

Seven discrete regions of X-ray emission are identified in NGC\,4449 
(Fig.\ \ref{fig5}); some are dominated by point sources while others by 
diffuse emission (Figs.\ \ref{fig4}a, b).  
Note that at the distance of NGC\,4449 ($\sim$5.4 Mpc), a ``point source"
could be as extended as 157 pc for the RHRI, and 780 pc for the PSPC.
Since X-ray point sources and diffuse
sources often have very different spectral characteristics 
(\cite{Williams95}), we may use the X-ray spectral properties and 
surface brightness to diagnose the nature of a source.  
We have extracted X-ray spectra from the PSPC data for these seven 
regions and for the entire galaxy.  
These spectra are displayed in Figs.\ \ref{fig7}a-h.  We have also used 
Raymond \& Smith's (1977) plasma emission models and Morrison 
\& McCammon's (1983) absorption characteristics to make spectral fits in 
order to derive X-ray luminosities.  For sources with insufficient counts for 
a spectral fit, we estimate the X-ray luminosity by adopting the Galactic 
H\,I column density log N$_{\rm H}$ = 20.2 (\cite{Fabbiano88}) for the 
absorption column and adopting plasma temperatures of well-observed 
interstellar objects with similar physical properties.
Below we describe these seven sources individually and discuss the 
integrated X-ray properties of NGC\,4449.  

Source 1 (hereafter S1) appears dominated by a point source 
(see Fig.\ \ref{fig4}b).  
Its relatively hard X-ray spectrum (Fig.\ \ref{fig7}a) is consistent with 
those commonly seen in X-ray binaries.  The number of counts is too low for 
a meaningful spectral fit.  Assuming a Raymond \& Smith plasma emission 
model and a plasma temperature of kT = 1 keV, we estimate an X-ray 
luminosity of $\sim6 \times 10^{38}$ ergs s$^{-1}$ for S1 in the 
0.1--2.4 keV band.

Source 2 (hereafter S2) is a known SNR, about 10 times brighter than the 
X-ray-brightest SNR in the LMC, N132D (\cite{Blair83}).  
The RHRI image of S2 shows a core with a FWHM of 6\farcs8, which is 
close to the instrumental FWHM (\cite{David96}), superimposed on
extended emission.  The best fit to the PSPC spectrum gives a plasma 
temperature of 1.0 keV and an absorption 
column density of log N$_{\rm H}$ = 20.6.  The absorption column is 
significantly higher than the Galactic H\,I column density, indicating
a log N$_{\rm H}$ = 20.4 within NGC\,4449 itself.  The X-ray luminosity 
of S2 is $\sim1\times 10^{39}$ ergs s$^{-1}$ in the 0.1--2.4 keV band.

Source 3 (hereafter S3) is coincident with the giant H\,II region CM\,16. 
The RHRI image in Fig.\ \ref{fig4}b shows a point-like source at S3; its FWHM 
$\sim$ 8$''$ is slightly more extended than the instrumental width.
The spectral energy distribution of S3 (Fig.\ \ref{fig7}c) is best fitted by
a Raymond \& Smith model with a plasma temperature of 0.4 keV.  This 
temperature is typical for the plasma emission seen in SNRs, 
superbubbles, and giant H\,II regions (\cite{Chu93}; \cite{Williams95}).  
Since a SNR at the distance of NGC\,4449 will appear 
unresolved in the RHRI image, and since S3 is more extended than the 
instrumental FWHM, it is possible that most of the X-ray flux of S3 
is emitted by hot, shocked interstellar gas in a luminous SNR and/or 
within a region $\sim$200 pc across.  The low intensity extensions into 
SGS1, SB3 and SB4, and along FIL28 seen in Fig.\ \ref{fig6}b\ will 
also contribute to the PSPC spectrum of S3.  
The best Raymond \& Smith model fit gives an H\,I column density of 
$1 \times 10^{20}$ cm$^{-2}$, indicating little foreground H\,I 
absorption within NGC\,4449.  The X-ray luminosity is $\sim8 \times 
10^{38}$ ergs s$^{-1}$ in the 0.1--2.4 keV band.

Source 4 (hereafter S4) has a very low surface brightness and extends
over a large area to the northwest of the main body of NGC\,4449.
The spectrum of S4 is clearly soft, and is softer than those of most 
known sources of diffuse emission.  The soft spectrum could be caused 
by the combination of a low plasma temperature and a low foreground 
absorption column density.  The soft spectrum and the extended 
distribution of S4 suggest that the X-ray emission originates from
hot interstellar gas. 
S4 has too few counts to warrant a spectral 
fit.  We have re-binned the PSPC spectrum, adopted the Galactic
H\,I column density as the foreground absorption column, and compared
the spectrum to those calculated using Raymond \& Smith models for 
plasma temperatures in the range 0.1--0.4 keV.  We find a 0.2 keV model
best represents the observed spectrum, and estimate an X-ray luminosity
of $7 \times 10^{38}$ erg s$^{-1}$ for S4.

Source 5 (hereafter S5) is located in the southern outskirts of NGC\,4449.
It is superimposed on a relatively quiescent region with no obvious
star formation activities.  The RHRI image shows an unresolved source, while 
the PSPC shows a region dominated by diffuse emission.
The spectrum of S5 is as soft as that of S4.  If the unresolved source is
a real point source, the absence of hard X-ray photons implies that the 
contribution of the point source is very soft, too. This may imply that the 
point source belongs to the class of supersoft X-ray sources.  
Most supersoft X-ray sources have the bulk of X-ray emission below 0.5 keV.  
Alternatively the point source is highly variable, and was very faint during
the PSPC observation. In this case we cannot further constrain the nature 
of the point source. 
Since the spectral properties of S5 are very similar to those of S4, and 
the PSPC emission is clearly extended, we think that S5 consists 
of extended, low-temperature plasma emission and possibly a supersoft 
point source.  Using a plasma temperature of 0.2 keV
and the Galactic foreground absorption, we derive an X-ray luminosity of
$\sim 6 \times 10^{38}$ erg s$^{-1}$ for S5 in the 0.1--2.4 keV band.

Source 6 (hereafter S6) appears extended both in the PSPC image and in 
the RHRI image.  It is located northeast of CM\,16 in a region without
prominent star formation activities.  The X-ray spectrum of S6 is soft,
suggesting a hot plasma emission.

Source 7 (hereafter S7) is a very weak point source northwest of NGC\,4449.
No optical counterpart can be identified in our images.  The PSPC spectrum 
is too noisy to provide useful information.  It is not clear if this 
source is associated with NGC\,4449 at all.

The integrated spectrum of NGC\,4449 is shown in Fig.\ \ref{fig7}h.  It is 
clearly a composite spectrum with different temperature components.  A single 
temperature component Raymond \& Smith model is plotted over the 
spectrum to demonstrate this effect.  Using this one-temperature
Raymond \& Smith model fit, we derive a crude estimate of the luminosity
of NGC\,4449, $3.5 \times 10^{39}$ erg s$^{-1}$ in the 0.1--2.4 keV band.

\subsection{Relationship between Hot and Warm Interstellar Components}

To analyze the large-scale diffuse X-ray emission of NGC\,4449,
we use the energy bands R1--R7  (\cite{Snowden92}) and made maps 
in the (R1+R2), (R4+R5), and (R6+R7) bands,
which are centered roughly at 1/4 keV, 3/4 keV, and 1.5 keV, 
respectively.  As shown in Fig.\ \ref{fig8}a--c, the images in these three 
energy bands have different characteristics.
In the 1.5 keV band, the emission is dominated by point sources
(S1, S2, S3 and S7).
In the 3/4 keV band, both point sources (S2, S3, and S7, but not S1) 
and diffuse sources (S4, S5, and S6) are visible.
In the 1/4 keV band, both point sources and diffuse sources are visible,
but the point sources become much less prominent.
The variation of NGC\,4449's X-ray morphology from hard to soft energy 
bands is similar to that observed in M101 (\cite{Snowden95}),
suggesting that the 1/4 keV band is dominated by diffuse emission from
hot interstellar gas.

The \ha\ image overlaid by X-ray contours of the PSPC data in 
Fig.\ \ref{fig6}a and \ha\ image overlaid by X-ray contours of the RHRI data 
in Fig.\ \ref{fig6}b can be used to relate the hot, X-ray-emitting 
interstellar gas to the cooler, \ha-emitting interstellar component.
The spatial resolution of the PSPC (30$''$, or 780 pc) 
prohibits unambiguous associations of faint diffuse X-ray emission 
with individual structures that are smaller than $\sim$ 1000 pc.
The RHRI has a higher resolution, but it also has a higher background
which makes the detection of faint diffuse emission difficult. 
Thus we are unable to unambiguously identify in NGC\,4449 X-ray-bright 
superbubbles similar to N44 and N51D in the LMC (\cite{Chu90}).  
The low level emission coinciding with SB3, SB4 and SB8 cannot be claimed 
as detections. Much longer RHRI integration is required to verify this 
weak emission.
Below we will discuss the hot interstellar gas associated with giant H\,II 
regions, supergiant shells, and large-scale \ha\ filamentary structures.

The giant H\,II region CM\,16 and adjacent superbubbles and filaments 
form a complex occupying an area more than 1000 pc across.  
The X-ray source S3 is associated with the CM\,16 complex.  As described
in \S4.1, S3 consists of a slightly resolved X-ray source with spectral
characteristics suggestive of hot plasma emission.
Hot, shocked interstellar gas is commonly seen in giant H\,II regions, 
for example, 30 Dor (\cite{Wang91}), NGC\,604 (\cite{Yang96}), 
and NGC\,5471 (\cite{Williams95}).  It is interesting to note that
the X-ray source S3 in CM\,16 not only has an X-ray luminosity within the 
observed range among these giant H\,II regions, but also has an emitting
volume similar to those of 30 Dor and NGC\,604, $\sim$200 pc in diameter.
We therefore conclude that the X-ray source S3 in CM\,16 is similar to
those seen in other giant H\,II regions.
The giant H\,II region CM\,39, on the other hand, does not show the typical
X-ray emission expected in giant H\,II regions.  CM\,39 is superimposed on 
diffuse X-ray emission; no discrete X-ray feature can be identified. 

Among the five supergiant shells identified in the \ha\ image of 
NGC\,4449, SGS1 is in a confusing region close to the giant H\,II 
region CM\,16. The RHRI may show some emission at the position of SGS1, but 
the surface brightness is much to low to claim a detection 
(see Fig.\ \ref{fig6}b).  SGS3, SGS4, and SGS5 are not detected.  Only SGS2 
is clearly visible in the PSPC image, corresponding to the source S4 
described in \S4.1.  

SGS2 appears to be morphologically similar to the supergiant
shell LMC2, as both supergiant shells appear to be ``blisters" blown 
by a ridge of active star formation out of the plane, but still confined 
by the H\,I disk.
However, SGS2 is different from LMC2 in three respects: (1) SGS2 is twice 
as large as LMC2, (2) the plasma temperature of SGS2 is much lower than that 
of LMC2, and (3) the X-ray luminosity of SGS2 is up to a factor of 10 higher 
than that of LMC2 (\cite{Points97}).  Thus, SGS2 appears to be the 
most energetic supergiant shell known!  

It is worth noting that the X-ray source S5 might be associated with
a not-well-defined supergiant shell.  The X-ray properties of S5 
imply the existence of an extended component emitted by plasma.
This X-ray emitting region is bounded by two long \ha\ filaments (FIL4 
and FIL5 in Fig.\ \ref{fig3}) in the northwestern and southwestern
quadrants.  These two long filaments may be the brightest parts of
a SGS.  Deeper \ha\ images are needed to confirm the existence of this
supergiant shell.

There are still X-ray emitting regions not bounded by obvious 
interstellar shell structures.
Most notably, bright diffuse X-ray emission exists along the ridge of
active star formation and extends outward into the adjacent quiescent 
regions.  The X-ray emission along the ridge is visible in both
the PSPC and the RHRI images.  This emission is particularly bright 
in the region between the X-ray sources S2 and S3, where the largest
number of chimney-like filaments are present.  These filaments 
might be physically associated with the X-ray emitting gas, 
for example, the region of FIL22, FIL23 and FIL3.  
These filaments and the diffuse X-ray emission might have be energized 
by the same massive star population.  
The X-ray emission region S6 is adjacent to but does not contain
star formation activities.  The lack of high concentration of massive
stars, shown below in \S5.1,  suggests that the hot gas is not 
energized locally and that the hot gas must have been transported 
from elsewhere.  It is likely that an outflow from the star formation 
region supplies this hot gas.  This mechanism might also be responsible 
for the hot gas in the SGS2 (X-ray source S4) and a possible SGS around 
the X-ray source S5.

\section{Discussion}

We have detected diffuse X-ray emission within the largest supergiant 
shell, within the active star forming regions, and in outflows from the 
star forming regions in NGC\,4449.  
The hot gas is most likely heated by massive stars via fast stellar
winds and supernova blasts.  
We have used an archival {\em Hubble Space Telescope (HST)} WFPC2 F606W 
(broad V) image of NGC\,4449 to examine in detail 
the distribution of massive stars.\footnote{The WFPC2 image allows us 
to identify OB associations as well as the brightest single stars.  
Since the brightest stars in the V band are supergiants evolved from 
massive stars, they can be used to trace the distribution of the 
massive stellar content in NGC\,4449.}
Below we relate the existence of hot interstellar gas to the underlying 
massive stellar population, present a quantitative analysis of the 
physical properties of the supergiant shell SGS2 and compare our detection 
of diffuse X-ray emission in NGC\,4449 to the results for other irregular 
galaxies.

\subsection{Hot Interstellar Gas and Massive Stars}

The massive stars and OB associations in the northern portion of
NGC\,4449 can be seen in the {\em HST} WFPC2 V image presented in 
Fig.\ \ref{fig9a}.  This region contains our X-ray sources S1, S2, S4, 
and S6.  Fig.\ \ref{fig9b} shows the RHRI contours overplotted over the 
{\em HST} image. 
S1 is likely an X-ray binary source, hence will not be
discussed further.  We only note that the RHRI indicates that S1 is 
multiple.  The other three sources all contain diffuse
emission.  These sources will be discussed individually. 

S2 contains a super-luminous SNR, which appears as a point source in
the RHRI image.  There is additional diffuse emission extending 
up to 14$''$ from the SNR in the NW and SE directions (see Fig.\ \ref{fig9b}).
Our ground-based \ha\ image shows that the SNR is within a bright
H\,II region.  The WFPC2 image shows that this SNR is in a large
stellar complex with a particularly high concentration of massive 
stars at 10-20$''$ SE of the SNR.  This correlation suggests that
the hot gas is heated locally by the massive stars.

S4 is a large region of diffuse X-ray emission that coincides with 
the supergiant shell SGS2.  SGS2 is bordered by a ridge of active star 
formation on the eastern side.  The WFPC2 image shows sparsely distributed 
massive stars within the boundary of SGS2 but a high concentration of 
massive stars along the eastern border, confirming the intense star 
formation activity.
The low surface density of massive stars within SGS2 casts doubt on 
the local production of the hot gas.  The star formation region
at the base of SGS2 thus becomes a likely candidate providing the
heating.  Indeed, the \ha\ image reveals multiple radial filaments
pointing from the star forming region into SGS2, reminiscent of
``chimneys" tracing the paths of energy transport (\cite{Heiles93}).  
The RHRI overlay image (Fig.\ \ref{fig9b}) shows diffuse emission extending 
along the northern part of the bar and north-east of CM\,16 where many 
\ha\ filaments point away from the bar into SGS2 (see Fig.\ \ref{fig2}). 
This further supports the identification of the radial filaments 
as chimneys which transport of hot gas upward from the star forming regions 
in the disk into SGS2. 
An additional confirmation comes from the echellograms published by HG90. 
The bright filament FIL3 (object 3 in HG90) appears as a 
high-velocity spur in their echellograms ``c, d, and e'', 
moving at +150 \kms\, with respect to the systemic velocity of 
NGC\,4449 at this position.

The diffuse X-ray source S6 is located in a region with faint featureless 
\ha\ emission. Fig.\ \ref{fig9b}  reveals only
few massive stars in this region, which are unlikely to be the powering
source of the hot gas.  The gas is likely transported to this position. 
Since this region is bordered by active
star formation regions on three sides and fragments of filaments
(FIL9, FIL10, and FIL38) on the other side, it is possible that
the hot gas originates from the star forming regions and is 
bounded by a shell structure.  

These three X-ray sources and the diffuse X-ray emission from the
active star formation ridge provide circumstantial evidence that
massive stars are responsible for heating the hot gas.  In the
active star formation region, the hot gas is heated locally.  In
quiescent regions adjacent to active star formation regions, 
hot gas can be heated in the active regions and transported 
to quiescent regions via blow-outs.

\subsection{Hot Gas Content of the Supergiant Shell SGS2}

We may use the PSPC observation to analyze the physical conditions
of the hot gas inside the supergiant shell SGS2.  
As described in \S4.1, the X-ray luminosity ($L_x$) of SGS2 in the 
0.1--2.4 keV band is $7\times10^{38}$ erg~s$^{-1}$, and the plasma 
temperature is estimated to be $kT \sim$ 0.2 keV.  Assuming that the 
X-ray emitting hot gas fills the interior of SGS2 (filling-factor of the 
hot gas $f = 1$), we may determine the rms electron density $n_e$ of 
the hot gas using the relation 

\noindent
$L_x~=~{n_e^2}~\Lambda_R(T)~V$,

\noindent 
where $\Lambda_R(T)$ is the emissivity in the 0.1--2.4 keV band and
$V$ is the volume.  For a temperature of 
$kT \sim$ 0.2 keV and a 30\% solar abundance, $\Lambda_R(T)$ 
is about $9.3 \times 10^{-24}$ erg cm$^3$ s$^{-1}$ for a Raymond \&
Smith's thin plasma emission model.  For a diameter of 1800 pc, the 
volume is $3\times10^{9}$ pc$^3$.  The rms electron density is 
calculated to be 0.03 cm$^{-3}$.  This density is higher than that 
derived for the supergiant shell LMC4, 0.01 cm$^{-3}$ (\cite{Bomans94}).  
Given the higher X-ray surface brightness of SGS2, 
its higher density is not too surprising.

With the density and volume described above, we derive a total
mass of $2\times10^6$ M$_\odot$, and a total thermal energy $E_{th}$ of 
$1.3\times10^{54}$ erg for the X-ray emitting plasma.  Note that
these values are upper limits for the mass and thermal energy.
If the X-ray emitting gas does not fill the entire interior of
the SGS ($f < 1$), the density 
would be higher, but the mass and thermal energy would be lower.

The cooling time scale for the hot gas inside SGS2 is about 
$2 \times 10^7$ yrs, assuming a cooling function of 
$2 \times 10^{-23} \times n_e^2$ ergs cm$^{-3}$ s$^{-1}$ and $f = 1$.  
The dynamical time scale for SGS2 is in the order of 
$5 \times 10^7$ yrs, assuming an expansion velocity of 20 km s$^{-1}$ and a 
radius of 2 kpc. Note that this timescale represents a rough upper limit 
for the dynamical age of SGS2, since it does not 
incorporate the acceleration due to expansion into a density gradient 
(e.g. \cite{MacLow89}).
The comparison between the cooling timescale and the dynamical timescale 
implies that energy input from the recent star formation event is needed 
to maintain the hot gas inside SGS2.

Future analysis of the energy budget based on detailed stellar
content and detailed nebular dynamics together with higher sensitivity and 
resolution X-ray data are needed to understand if SGS2 is indeed powered by 
the star forming region at its base.

\subsection{Comparison to Other Starforming Dwarf Galaxies}

NGC 4449 is the first normal star forming irregular galaxy besides the 
LMC in which diffuse X-ray emission from hot gas is detected.  
The origins of diffuse X-ray emission in NGC\,4449 are similar to those
in the LMC (Chu 1996), such as giant HII region, supergiant shells, 
superbubble, and quiescent regions with no recent star formation 
activities.  The supergiant shell SGS2 in NGG\,4449 is more extreme in 
its properties than all LMC supergiant shells.  We find evidence indicating 
that energy is being pumped into SGS2 in NGC\,4449; simliar energy 
input is seen in the supergiant shell LMC\,2 (\cite{Points97}). 
The mechanisms for producing diffuse X-ray emission appear to be 
similar for NGC\,4449 and the LMC.

We can also compare NGC\,4449 to more active star forming irregular galaxies. 
X-ray observations of four starbursting irregular galaxies have been 
reported: NGC\,1569, NGC\,1705, NGC\,5253, and UGC\,6456.  
The amorphous irregular galaxies NGC\,5253 (\cite{Martin95}) and 
NGC\,1705 (\cite{Hensler96}) show diffuse X-ray 
emission near the starburst region where super starclusters have been 
detected (\cite{Meurer95}); furthermore the diffuse emission 
in NGC\,1705 seems to be bounded by \ha\ shells.  
The blue compact dwarf galaxies NGC\,1569 (\cite{Heckman95}) and 
UGC\,6456 (\cite{Papaderos94}) show diffuse X-ray emission at large 
distances from the starburst regions, which has been interpreted as 
outflows.  Interestingly, two super starclusters are found in the 
starburst core of NGC\,1569 (\cite{OConnell94}).

All of these four galaxies show hot gas produced predominatly in their 
starburst regions.  This is similar to what we see in NGC\,4449 and the 
LMC, although star formation in the latter two galaxies is more spreaded 
over the galaxy and at an overall lower level.  
The two low-mass galaxies with most intense star formation 
(NGC\,1569 and UGC\,6456) show large-scale outflow of hot gas.  
This is similar what we see in SGS2 and the diffuse X-ray sources (such as 
S6) in NGC\,4449. 
This indicates that outflows from intense star formation regions may be 
common in star forming irregular galaxies. 

Our analysis of NGC\,4449 indicates that its production of hot gas is 
similar to that of the LMC and other active star forming irregular 
galaxies. However, we still do not know why the SMC does not 
show appreachiable diffuse X-ray emission despite its moderate star 
formation activity. 
The irregular galaxy Ho\,IX might have similar X-ray properties as the 
SMC.  Diffuse X-ray emission has reported in 
(\cite{Miller95}), however, the hard X-ray spectrum of this source 
is more consistent with that of an X-ray binary; no soft X-ray component 
is present to indicate the existence of hot interstellar gas.  
It is possible that a certain level of star formation activity is required  
to produce the hot gas responsible for the diffuse X-ray emission.
A larger sample of irregular galaxies with different levels of star 
formation activity need to be studied in order to determine whether 
and how star formation activity governs the level of diffuse X-ray emission 
in irregular galaxies.

\acknowledgments
We thank D. Psaltis for discussions on X-ray binaries, 
M. Wrigge for discussions on subtracting PSPC's point sources of an extended 
background, and K. Weis for comments on the manuscript.  This research 
made use of the NASA's Astrophysics Data System Abstract Service and 
the NASA/IPAC Extragalactic Database (NED) which is operated by the Jet 
Propulsion Laboratory, Caltech, under contract with NASA. 
DJB thanks the Alexander von Humboldt-Gesellschaft for partial
support through the Feodor Lynen fellowship program.
YHC is supported by NASA grants NAG 5-1900 (ROSAT) and
NAG 5-3246 (LTSA).
UH acknowledges the support by the Sonderforschungsbereich 375 of 
the Deutsche Forschungsgemeinschaft.

\newpage
\appendix

\section{Appendix}
To ease the discussion of H$\alpha$ features in this paper, we have
identified and listed in Tables \ref{tabA1} and \ref{tabA2} the principal 
diffuse \ha\ features in NGC\,4449.  Table \ref{tabA1} lists the supergiant 
shells and superbubbles of NGC\,4449.  For each shell structure the central
coordinates were determined, using the STSDAS astrometry package and 
the digital sky survey for the astrometric reduction.  We also measured
the angular sizes of the shells and converted them to linear sized
assuming a distance of 5.4 Mpc.  Brief comments are also given.
Table \ref{tabA2} lists the principal H$\alpha$ filaments in NGC\,4449. 
The list is not complete, especially in the very crowded central region.
The coordinates of the brightest parts or the centers of the filaments 
are given.

\newpage
\centerline{\large \bf Figure Captions}

\figurenum{1}
\figcaption[fig1.ps]{Red continuum image of NGC\,4449, taken with the Calar 
Alto 3.5m telescope. \label{fig1}}

\figurenum{2}
\figcaption[fig2.ps]{Continuum-subtracted \ha\ image of NGC\,4449. 
Shells and filaments are prevalent in the main body of NGC\,4449.
Faint shells and filaments extend to large distances from the
main body. \label{fig2}}
		
\figurenum{3}      
\figcaption[fig3.ps]{Continuum-subtracted \ha\ image of NGC\,4449 with the 
principal shells and filaments marked. \label{fig3}}

\figurenum{4}
\figcaption[fig4a.ps,fig4b.ps]{(a) Smoothed {\em ROSAT} PSPC image of 
NGC\,4449 in the 0.1--2.4 keV energy band, displayed with the same image 
scale as the optical images in Figs.\ 1 and 2.
(b) Smoothed {\em ROSAT} HRI image of NGC\,4449, displayed with the 
same image scale as the images in Figs.\ 1, 2, and 4a. \label{fig4}}

\figurenum{5}
\figcaption[fig5.ps]{Broad-band (0.1--2.4 keV), un-smoothed PSPC image of 
NGC\,4449.  The seven source regions are marked.  The large un-numbered 
circle encloses the region for the entire NGC\,4449 galaxy. \label{fig5}}

\figurenum{6}
\figcaption[fig6a.ps,fig6b.ps]{(a) Continuum-subtracted \ha\ image of 
NGC\,4449 overlaid by X-ray contours derived from the broad-band 
(0.1--2.4 keV) {\em ROSAT} PSPC image.
The same continuum-subtracted \ha\ image of NGC\,4449 overlaid by
the {\em ROSAT} HRI image. \label{fig6}}

\figurenum{7}
\figcaption[fig7.ps]{{\em ROSAT} PSPC energy distributions of the seven X-ray 
sources and of the entire NGC\,4449 galaxy.  Sources 2 and 3 have 
sufficient counts for spectral fits.  The best-fit Raymond \& Smith
models are over-plotted.  The one-temperature component Raymond \&
Smith model fit for the entire galaxy is clearly an 
over-simplification.  This fit is used for only a crude estimate 
of the luminosity of NGC\,4449. \label{fig7}}

\figurenum{8}
\figcaption[fig8a.ps,fig8b.ps fig8c.ps]{{\em ROSAT} PSPC images of NGC\,4449 
in (a) the (R1+R2) band centered at 1/4 keV, (b) the (R4+R5) band centered 
at 3/4 keV, and (c) the (R6+R7) band centered at 1.5 keV. \label{fig8}}

\figurenum{9a}
\figcaption[fig9a.ps]{{\em HST} WFPC2 image of NGC\,4449, taken through the 
F606W (broad V) filter. \label{fig9a}}

\figurenum{9b}
\figcaption[fig9b.ps]{{\em HST} WFPC2 image of NGC\,4449, taken through 
the F606W (broad V) filter. The overlayed contours are X-ray emission taken 
from the RHRI image. \label{fig9b}}

\begin{deluxetable}{rccccc}
\tablenum{1}
\label{tab1}
\tablewidth{0pt}
\tablecaption{Basic X-ray Data of the Sources}
\tablehead{
\colhead{Region} & \colhead{Counts} & 
\colhead{log N$_H$} & \colhead{kT}  & 
\colhead{L$_X$(0.1-2.4 keV)} & \colhead{L$_X$(0.5-2.4 keV)} \\
\colhead{} & \colhead{} & 
\colhead{[cm$^{-2}$]} & \colhead{[keV]} & 
\colhead{[10$^{38}$ erg s$^{-1}$]} & \colhead{[10$^{38}$ erg s$^{-1}$]}
}
\startdata 
1    &  98 $\pm$ 17 & 20.2 & 1.0 &  6 &  2 \nl
2    & 195 $\pm$ 21 & 20.6$^a$ & 1.0$^a$ & 13 &  8 \nl
3    & 181 $\pm$ 20 & 20.2$^a$ & 0.4$^a$ &  8 &  5 \nl
4    & 116 $\pm$ 20 & 20.2 & 0.2 &  7 &  2 \nl
5    &  98 $\pm$ 18 & 20.2 & 0.2 &  6 &  2 \nl
6    & 101 $\pm$ 17 & 20.2 & 0.7 &  5 &  4 \nl
7    &  14 $\pm$ 13 & 20.2 & 0.7 &  1 &  1 \nl
All$^b$ & 947 $\pm$ 55 & 20.2 & 0.6 & 35  & 25 \nl
\enddata
\tablenotetext{a}{The log N$_H$ and kT values are result from a spectral 
spectra fit.}
\tablenotetext{b}{Integrated over the entire NGC\,4449 galaxy.}
\end{deluxetable}

\begin{deluxetable}{lcccl}
\tablenum{A1}
\label{tabA1}
\tablewidth{0pt}
\tablecaption{List of the Supergiant Shells and Superbubbles in NGC\,4449}
\tablehead{
\colhead{Source} & \colhead{RA} & 
\colhead{DEC} & \colhead{Diameter} & 
\colhead{Comment} \\
\colhead{} & \colhead{(2000.0)} &
\colhead{(2000.0)} & \colhead{[pc]} &
\colhead{} 
}
\startdata
SGS1 & 12 28 07.8 & 44 04 50  & 520 & structure 1 in HG90, expanding \nl 
SGS2 & 12 28 06.4 & 44 06 25  & 1830 & structure 5 \& 6 in HG92 \nl 
SGS3 & 12 27 56.6 & 44 05 40  & 500 &    \nl 
SGS4 & 12 28 05.5 & 44 05 00  & 1060 & diffuse  \nl 
SGS5 & 12 28 17.7 & 44 05 10  & 1295 & diffuse, uncertain  \nl 
SB1 &  12 28 20.9 & 44 06 11  & 360 &    \nl 
SB2 &  12 28 09.7 & 44 06 54  & 340 &    \nl 
SB3 &  12 28 10.1 & 44 05 09  & 235 & discussed in H94; part of CM16? \nl 
SB4 &  12 28 10.4 & 44 05 06  & 475 & structure 2 in HG90; part of CM16 ?, SGS6 ? \nl
SB5 &  12 28 13.8 & 44 05 02  & 390 &    \nl 
SB6 &  12 28 05.4 & 44 04 16  & 255 & attached filament \nl 
SB7 &  12 28 06.6 & 44 03 41  & 255 &    \nl 
SB8 &  12 28 06.7 & 44 04 23  & 235 &    \nl 
SB9 &  12 28 13.6 & 44 04 33  & 465 & SGS7 ?   \nl
\enddata 
\end{deluxetable}

\clearpage
\begin{deluxetable}{lccl}			      
\tablenum{A2}
\label{tabA2}
\tablewidth{0pt}
\tablecaption{List of the Principal H$\alpha$ Filaments in NGC\,4449.}  
\tablehead{
\colhead{Source} & \colhead{RA} & 
\colhead{DEC} & \colhead{Comment} \\
\colhead{}   & \colhead{(2000.0)} &
\colhead{(2000.0)} & \colhead{}
}
\startdata
FIL1 &	   12 28 13.2 & 44 06 34   & partly structure 4 in HG90 \nl
FIL2 &	   12 28 07.7 & 44 05 41   &    \nl
FIL3 &	   12 28 09.6 & 44 05 58   & partly structure 3 in HG90 \nl
FIL4 &	   12 28 08.0 & 44 04 20   &    \nl
FIL5 &	   12 28 08.2 & 44 03 42   &    \nl
FIL6 &	   12 28 11.7 & 44 05 04   & interlocking shells ?   \nl
FIL7 &	   12 28 15.2 & 44 05 52   &    \nl
FIL8 &	   12 28 14.4 & 44 05 59   & group of filaments   \nl
FIL9 &	   12 28 18.0 & 44 05 38   &    \nl
FIL10 &	   12 28 19.0 & 44 06 14   &    \nl
FIL11 &	   12 28 17.0 & 44 06 29   &    \nl
FIL12 &	   12 28 17.8 & 44 06 59   & superbubble ?   \nl
FIL13 &	   12 28 14.4 & 44 06 52   &    \nl
FIL14 &	   12 28 14.9 & 44 06 59   &    \nl
FIL15 &	   12 28 12.4 & 44 07 11   & `intersecting' filaments   \nl
FIL16 &    12 28 11.6 & 44 07 04   &    \nl
FIL17 &	   12 28 10.1 & 44 07 06   &    \nl
FIL18 &	   12 28 08.1 & 44 06 51   &    \nl
FIL19 &	   12 28 09.1 & 44 06 30   &    \nl
FIL20 &	   12 28 12.1 & 44 06 36   &    \nl
FIL21 &	   12 28 12.1 & 44 06 24   & superbubble ?   \nl
FIL22 &	   12 28 10.1 & 44 06 19   & `intersecting' filaments   \nl
FIL23 &	   12 28 11.5 & 44 06 08   &    \nl
FIL24 &	   12 28 10.9 & 44 05 59   & complex web with FIL3 and 25   \nl
FIL25 &	   12 28 10.3 & 44 05 45   & open shell ? \nl
FIL26 &	   12 28 10.3 & 44 05 38   & open shell ? \nl
FIL27 &	   12 28 09.8 & 44 05 31   &    \nl
FIL28 &	   12 28 09.1 & 44 05 28   & part of CM16 ?   \nl
FIL29 &	   12 28 08.2 & 44 05 16   & open shell?; part of CM16 ?   \nl
FIL30 &	   12 28 12.1 & 44 05 21   & interlocking shells ?   \nl
FIL31 &	   12 28 12.3 & 44 04 52   &    \nl
FIL32 &	   12 28 15.1 & 44 04 53   & group of filaments   \nl
FIL33 &	   12 28 14.0 & 44 06 17   & group of filaments   \nl
FIL34 &	   12 28 12.4 & 44 06 51   &    \nl
FIL35 &	   12 28 15.0 & 44 06 31   & multiple filaments   \nl
FIL36 &	   12 28 17.1 & 44 07 13   &    \nl
FIL37 &	   12 28 07.2 & 44 05 53   &    \nl
FIL38 &	   12 28 17.8 & 44 05 30   &    \nl
\enddata     	      
\end{deluxetable}  	      
		      
\end{document}